# A Dedicated Modelling Scheme for Nonclassical Optical Response from the Nanosphere-on-Mirror Structure

Xiaotian Yan, Christos Tserkezis, N. Asger Mortensen, Guy A. E. Vandenbosch, *Fellow IEEE* and Xuezhi Zheng, *Member, IEEE*

*Abstract*—Within the framework of the T-matrix method, we present a modeling tool that predicts the optical response from the Nanosphere-on-Mirror (NSoM) construct. The nonclassical effects in metals are accounted for by the nonlocal hydrodynamic Drude model (NLHDM) or the surface response model (SRM). Two essential elements in the T-matrix method, i.e., the T-matrix of the sphere and the R matrix accounting for the effects of the mirror, have been fully upgraded to include longitudinal waves for the NLHDM and the augmented interface conditions for the SRM. The proposed tool is quantitatively validated both in the near and the far field by an in-house developed BEM solver for the NLHDM where the gap between the sphere and the mirror is as small as 1 nm. Two physical checks are performed, where the results from the classical local response model are compared with the ones from the NLHDM and the SRM. The observed shifts in resonances and reduced field enhancements in the gap region agree well with previous physical findings. The proposed tool may not only serve as a reference tool for other numerical methods, but also provides an ideal platform for investigating nonclassical optical processes in the NSoM, hence paving a semi-analytical way to understand the extreme optics at very small scales.

X. Yan is grateful for the China Scholarship Council (Grant No. 202208320006), China. X. Zheng, and G. A. E. Vandenbosch are grateful for the C1 project (C14/19/083), the IDN project (IDN/20/014), the small infrastructure grant (KA/20/019) from KU Leuven and for G090017N and G088822N from the Research Foundation of Flanders (FWO). X. Zheng would like to also thank the support from the IEEE Antennas and Propagation Society Postdoctoral Fellowship, and for the FWO long travel grant, FWO V408823N. N. A. Mortensen is a VILLUM investigator supported by VILLUM Fonden (Grant No. 16498). The Center for Polariton-driven Light – Matter Interactions (POLIMA) is funded by the Danish National Research Foundation (Project No. DNRF165). (Corresponding Author: Xuezhi Zheng).
X. Yan, X. Zheng, and G. A. E. Vandenbosch are affiliated with the WaveCore Division, Department of Electrical Engineering (ESAT), KU Leuven, B-3001, Leuven, Belgium.
X. Zheng, C. Tserkezis and N. A. Mortensen are affiliated with POLIMA – Center for Polariton-driven Light-Matter Interactions, University of Southern Denmark, 5230 Odense, Denmark.

Color versions of one or more of the figures in this article are available online at http://ieeexplore.ieee.org

## I. INTRODUCTION

THE Nanoparticle-on-Mirror (NPoM) structure [1] consists of a metal nanoparticle (NP) positioned on top of a mirror with, e.g., a self-assembled molecular monolayer (a gap layer) in between. The thickness of the monolayer fixes the cavity gap size at the deep-nanometric scales, i.e., from a fraction of one nanometer (nm) to a few nms. As a result, a strong field enhancement (exceeding hundred-fold with respect to the magnitude of the incident field) is formed in the gap region of the NPoM, which makes the NPoM an ideal platform for many ground-breaking applications, e.g., perfect absorbing artificial medium [2], rapid nanoscopic imaging [3], up-converting mid-IR light to the optical band [4], [5], to name a few.

Modelling the interaction of light with the NPoM plays an important role in theoretically understanding how the light is molded by the nanocavity. This is often done within the scope of the local response model (LRM) where the optical response of the metals constituting the NP and the mirror is described by a frequency dependent dielectric function of the bulk material, so that many computational electromagnetics (CEM) techniques, e.g., the Finite Difference Time Domain (FDTD) method [6], the Finite Element Method (FEM) [7], the Discontinuous Galerkin (DG) Method [8], the Boundary Element Method (BEM) [9], and the Volumetric Method of Moments (MoM) [10], can be (re-)employed. Empowered by many post-processing techniques, e.g., mode [11]–[13] and symmetry analysis [14], [15], the mode structure, the near field and the far field [4], [16] of the NPoM have been thoroughly investigated.

However, due to the deep-nanometric nature of the cavity, the non-classical effects in metals, which transcend the LRM, play a non-negligible role in shaping the optical response of the NPoM [17]–[19]. For these effects, many semiclassical models are proposed. Amongst these models, two important categories are: the nonlocal hydrodynamic Drude model (NLHDM) [20]–[25] which employs a fluidic picture to account for the finite compressibility of electron gas, and its extensions to include diffusion [22], i.e., the generalized nonlocal optical response (GNOR), and to include the electron spill-out, i.e., the self-consistent hydrodynamic model (SC-HDM) [23]; and, most recently, the surface response model (SRM) [26]–[29] which lumps the complicated light-matter interaction in the transition region around, e.g., a metal-vacuum interface, by a set of quantum corrected boundary conditions. These models successfully predict, e.g., spectral shifts [22], [26], [30]–[32], reduced near field enhancement [33], to name a few.

As a result, conventional CEM algorithms for classical electrodynamics must be systematically upgraded to cope with the challenges posed by these new physical models. Firstly, for canonical geometries, planar layers, cylinders, and spheres, can be (semi-)analytically analyzed within the framework of the T-matrix (or the S-matrix) algorithm for the NHDLM [31], [34]–[41] and the SRM [42]. Further, the differential equation (DE-) based methods, e.g., FDTD, FEM, DG-FEM, have been readily applied to study the non-classical effects for arbitrary nanotopologies, for the NLHDM [22], [39], [43]–[45], and for the SRM [27]. Lastly, the integral equation (IE-) based methods, e.g., BEM and V-MoM, have been tailored to study NPs both in homogeneous space (for the NLHDM [46]–[50] and for the SRM [51]) and on layers (for the NLHDM [52]).

Since the (semi-)analytical approach does not only provides an efficient way for solving the problem, but also always carries a vast amount of physical information, it is deemed as an essential element in physics and holds an irreplaceable position among the three aforesaid computational approaches. Motivated by this, in this work, we intend to extend the T-matrix algorithm for a single nanosphere (NS) to include the effects of the underneath mirror, so that the optical response of the NSoM structure can be computed by the T-matrix algorithm. We note that this has already been done for the LRM [53], and, for the NLHDM, has been done for a nonlocal particle on a *dielectric* substrate [54], [55]. Also, for the NLHDM, in [56], although a NSoM structure is considered, the mirror is treated as a perfect conductor, implying that the nonclassical effects are ignored in the mirror.

In this work, we propose a computational tool that considers the nonclassical effects (which can be either described by the NLHDM or the SRM) in both the NS and the mirror. In more details, the NS can be a concentric shell whose layers can be nonclassical metals and (isotropic) dielectrics, and the mirror can be a planarly stratified structure whose layers can be nonclassical metals and (isotropic and uniaxial) dielectrics. This work is organized as follows. Section II briefly reviews the NLHDM and the SRM where the main equations of the two models are summarized. Then, the main equation behind the computation is discussed. It is pointed out that the T-matrix (that describes the input – output relation for the NS) and the R-matrix (that covers the effects of the mirror) in the equation are the two elements to be upgraded. In Section III, we expand the spherical waves (SWs) radiated by the NS (in the top layer) in terms of plane waves (PWs), trace the multiple reflection of the PWs through the layers, collect reflected, scattered, and transmitted PWs in the layers, and expand the reflected PWs in the top layer by SWs. By such a four-step procedure, not only the R-matrix is constructed, but also the reflected, scattered, and transmitted SWs in all layers are obtained. Then, similar to the approach in our previous work [41], we find the S-matrix for a planar or a spherical interface. Since the NLHDM has been well covered in [41], the procedures for the SRM are deliberated. Lastly, in Section IV, we compare the results from the proposed tool with the ones from an in-house developed BEM solver [47], [52]. A good agreement is demonstrated. Besides, two physical checks are done for the NLHDM and the SRM where the observed physics are well in line with previously reported results [26], [33].

## II. THEORY

In this section, we first give a quick overview on the key elements of the NLHDM including GNOR [20], [22], [24] and the SRM [26], [27]. For more information on the models, we refer the readers to two most recent reviews [57], [58]. To conclude the section, we illustrate the main equation behind the proposed tool, and discuss the impact of the two models on the implementation of the tool.

In the work, we assume the $e^{-i\omega t}$ time dependency with $\omega$ being the angular frequency (accordingly $k_0$ being the vacuum wavenumber). For the sake of conciseness, $e^{-i\omega t}$ will be suppressed. To be complete, we list the Maxwell equations,

$$\nabla \times \mathbf{E}(\mathbf{r},\omega) = i\omega\mu_0 \mathbf{H}(\mathbf{r},\omega), \qquad (1)$$

$$\nabla \times \mathbf{H}(\mathbf{r},\omega) = \mathbf{J}(\mathbf{r},\omega) - i\omega \mathbf{D}(\mathbf{r},\omega). \qquad (2)$$

In the above two equations, $\mathbf{E}$, $\mathbf{H}$, $\mathbf{J}$ and $\mathbf{D}$ are the electric, the magnetic, the source current, and the electric displacement fields. $\mathbf{r}$ is a spatial point and $\mu_0$ is the vacuum permeability. Also, we assume that the material is non-magnetic (as shown by the vacuum permeability $\mu_0$ in Eq. (1)). Lastly, the SI units are used in the work.

### A. Nonlocal Hydrodynamic Drude Model (NLHDM)

The NLHDM and its extension, that is, GNOR, treat the free electron gas in a metal as an electron fluid, trace the motion and the force-balance of a *fluid particle*, i.e., a volume being *locally* seen as a *uniform* electron gas, and describes the dynamics, i.e., convection and diffusion, of the particle by an additional partial differential equation (PDE) to the Maxwell equations,

$$\xi^2 \nabla \left( \nabla \cdot \mathbf{P}_f(\mathbf{r}) \right) + \mathbf{P}_f(\mathbf{r}) = -\varepsilon_0 \frac{\omega_p^2}{\omega(\omega + i\gamma)} \mathbf{E}(\mathbf{r}). \qquad (3)$$

In Eq. (3), $\omega_p$ is the plasma frequency. $\gamma$ is the damping rate. $\mathbf{P}_f(\mathbf{r})$ and $\mathbf{E}(\mathbf{r})$ are the free-electron polarization current and the electric field at a spatial point $\mathbf{r}$. $\mathbf{P}_f(\mathbf{r})$ enters the Maxwell equations in Eq. (1) and Eq. (2) via the electric displacement field,

$$\mathbf{D}(\mathbf{r}) = \varepsilon_0 \varepsilon_{bd} \mathbf{E}(\mathbf{r}) + \mathbf{P}_f(\mathbf{r}). \qquad (4)$$

Here, $\varepsilon_{bd}$ is the bound-electron permittivity (see the definition in, e.g., [47]). Lastly, $\xi$ is, when only considering convection,

$$\xi^2(\omega) = \frac{\beta^2}{\omega(\omega + i\gamma)}, \qquad (5)$$

when considering both convection and diffusion [22], [45],

$$\xi^2(\omega) = \frac{\beta^2}{\omega(\omega+i\gamma)} + \frac{D}{i\omega}. \quad (6)$$

In the above, $\beta$ is a quantity which is related with the Fermi velocity, i.e., $\beta^2 = 3/5\ v_F^2$, in the high-frequency limit, and is closely related with the finite compressibility of the electron gas [58]. $D$ is the diffusion constant, i.e., a phenomenological parameter, which lumps possible microscopic processes, e.g., non-specular scattering at metal surfaces, surface enhanced Landau damping, to name a few. The physical model in Eq. (3) underlines the demolition of the concept of the surface charge in macroscopic EM. The charge induced by an external optical perturbation cannot stay on the boundary of the metal, must be "broadened" and occupy a finite volume. Its impact on the optical response of the NSoM will be seen in Section VI.B.

The extra PDE (besides the Maxwell equations) in Eq. (3) requires additional boundary conditions (ABCs) beyond the conventional BCs at material interfaces,

$$\mathbf{n} \times (\mathbf{E}_2 - \mathbf{E}_1) = \mathbf{0}, \quad (7)$$

$$\mathbf{n} \times (\mathbf{H}_2 - \mathbf{H}_1) = \mathbf{0}. \quad (8)$$

For a metal – dielectric interface, we *assume* the ABC as,

$$\mathbf{n} \cdot \mathbf{P}_f = 0. \quad (9)$$

Eq. (9) marks the termination of the free electron polarization current at the metal boundary, by saying that no electrons can escape from the metal. For a metal – metal interface, we *say*,

$$\mathbf{n} \cdot \mathbf{P}_{1,f} = \mathbf{n} \cdot \mathbf{P}_{2,f}, \quad (10)$$

$$\frac{\beta_1^2}{\omega_{1,p}^2}\varepsilon_{1,bd}\nabla \cdot \mathbf{E}_1 = \frac{\beta_2^2}{\omega_{2,p}^2}\varepsilon_{2,bd}\nabla \cdot \mathbf{E}_2. \quad (11)$$

Eq. (10) and Eq. (11) stem from the requirement of continuous normal component of the energy current density (see chapter 2 in [20]). It is underlined that Eq. (9) – Eq. (11) are *selected* in an ad-hoc manner. In the above, **n** is the boundary normal.

### B. Surface Response Model (SRM)

The SRM focuses on a "transition" (selvedge) region [59] between a metal and, e.g., vacuum. Instead of treating the charge in the region induced by an external EM wave as a "strict" surface one, the SRM considers the polarizable dipole moments of the induced charge distribution at the metal – vacuum interface as well. This leads to the quantum-corrected boundary conditions (QC-BCs) [26], [27], [58]. Related to the proposed scheme are the two BCs regarding the tangential components of the **E**- and **H**- fields,

$$\mathbf{E}_2^\parallel - \mathbf{E}_1^\parallel = -d_\perp \cdot \nabla_\parallel \left( E_2^\perp - E_1^\perp \right), \quad (12)$$

$$\mathbf{H}_2^\parallel - \mathbf{H}_1^\parallel = -i\omega d_\parallel \cdot \mathbf{n} \times \left( \mathbf{D}_2^\parallel - \mathbf{D}_1^\parallel \right). \quad (13)$$

In the above, $d_\perp$ and $d_\parallel$ are Feibelman parameters [60]. They are very related to the dipole moments of the induced charge distribution normal to and along the boundary of the metal and can be determined from a Time-Dependent Density Functional Theory (TD-DFT) calculation [42], or even from LRM and spatially varying equilibrium electron density [29]. Clearly, when $d_\perp$ and $d_\parallel$ are set to zero, the conditions in Eq. (12) and Eq. (13) reduce to the conventional BCs. Here, the subscripts "1" and "2" mark the physical quantities related to the inner region and the outer region of the boundary. The superscripts "⊥" and "∥" refer the directions normal to and in parallel with the boundary. **n** is the boundary normal.

### C. Problem Statement and Main Equation

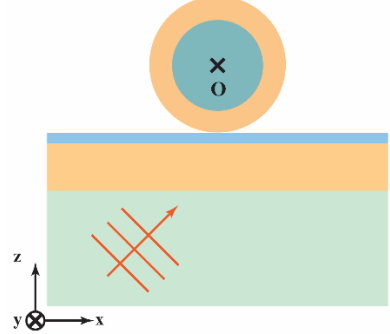

Fig. 1. Illustration of the NSoM structure. In the figure, the "sphere" is a core-shell structure where the core and the shell are made of dielectrics and metals; and the "mirror" composes of three layers, i.e., a thin gap marked by the blue color, a thin metal film marked by the yellow color and a dielectric substrate marked by the green color, which have an infinite extension in the *x-y* plane. A coordinate system is attached to the structure (see the bottom-left) and the origin of the coordinate system is fixed at the center of the "sphere".

The proposed tool is dedicated to modelling the interaction of light with the NSoM. From a computational point of view, the following abstraction has been made (see Fig. 1). First of all, the NS is not necessary to be a homogeneous sphere but can be in a concentric shell topology, and the materials filling the NS can be (isotropic) dielectrics and metals. Second, the NS is placed in the top layer (which is always assumed to be filled by an isotropic dielectric) of a planar multilayer structure. And here after, we refer to the layers underneath the NS as the "mirror". The layers can be filled by (isotropic and/or uniaxial which models, for example, graphene) dielectrics and metals. Lastly, the external excitations are PWs coming in from the top layer or the bottom layer, which is in line with what is commonly used in experimental setups. As a remark, the origin of the coordinate system is set at the center of the NS.

The key relation behind the modelling is known from the T-matrix algorithm [53],

$$\begin{pmatrix} \mathbf{a}^s \\ \mathbf{b}^s \end{pmatrix} = \mathbf{T} \cdot \begin{pmatrix} \mathbf{a}^e \\ \mathbf{b}^e \end{pmatrix}. \quad (14)$$

In Eq. (14), **T** is the transition matrix. The **T**-matrix links the expansion coefficients of the **total** incident field with the ones of the **direct** scattered field, and only depends on the geometry of the NS and materials filling the NS but is independent of the underneath layers. In detail, the expansions of the **total** incident field and the **direct** scattered field are,

$$\mathbf{E}^e(\mathbf{r}) = \sum_{nm} \left[ \mathbf{M}_{nm}^e(\mathbf{r}) \cdot a_{nm}^e + \mathbf{N}_{nm}^e(\mathbf{r}) \cdot b_{nm}^e \right], \quad (15)$$

$$\mathbf{E}^s(\mathbf{r}) = \sum_{nm} \left[ \mathbf{M}_{nm}^s(\mathbf{r}) \cdot a_{nm}^s + \mathbf{N}_{nm}^s(\mathbf{r}) \cdot b_{nm}^s \right]. \quad (16)$$

In Eq. (15) and Eq. (16), the superscripts "*e*" and "*s*" refer to

the expansions of the **total** incident field and the **direct** scattered field by the so-called *standing* and *radiating* SWs. This corresponds to the use of spherical Bessel or Hankel function in the **M** and **N** functions in Eq. (16) and Eq. (15) (see the detailed forms of the **M** and **N** functions in the Chapter 7 in [61]). The subscripts $nm$ refer to the azimuthal and magnetic quantum number, and they constrain the angular variations of the **M** and the **N** functions. Further, $n$ is a positive integer and, for a given $n$, $m$ is an integer between $-n$ and $n$. In this work, we always consider column vectors whose elements are $a_{nm}^j$ and $b_{nm}^j$, i.e.,

$$\mathbf{a}^j = \{a_{1,-1}^j,\ldots,a_{l,l}^j\}^T, \quad \mathbf{b}^j = \{b_{1,-1}^j,\ldots,b_{l,l}^j\}^T. \quad (17)$$

In Eq. (17), the number of elements in each column vector is $N = l^2$ and "T" marks the matrix transpose.

Further, the **total** incident field includes two contributions one from an external excitation, i.e., $\mathbf{E}^i$, the other from the **reflected** scattered field, i.e., $\mathbf{E}^r$, as the result of the interaction of the **direct** scattered field with the "mirror". Both fields must be expanded in terms of *standing* SWs,

$$\mathbf{E}^i(\mathbf{r}) = \sum_{nm}\left[\mathbf{M}_{nm}^e(\mathbf{r})\cdot a_{nm}^i + \mathbf{N}_{nm}^e(\mathbf{r})\cdot b_{nm}^i\right], \quad (18)$$

$$\mathbf{E}^r(\mathbf{r}) = \sum_{nm}\left[\mathbf{M}_{nm}^e(\mathbf{r})\cdot a_{nm}^r + \mathbf{N}_{nm}^e(\mathbf{r})\cdot b_{nm}^r\right]. \quad (19)$$

The sum of the expansion coefficients in Eq. (18) and the ones in Eq. (19) gives the ones of the **total** incident field,

$$\begin{pmatrix}\mathbf{a}^e\\\mathbf{b}^e\end{pmatrix} = \begin{pmatrix}\mathbf{a}^i\\\mathbf{b}^i\end{pmatrix} + \begin{pmatrix}\mathbf{a}^r\\\mathbf{b}^r\end{pmatrix}. \quad (20)$$

In the above equation, $\mathbf{a}^i$, $\mathbf{b}^i$ and $\mathbf{a}^r$, $\mathbf{b}^r$ are column vectors of the expansion coefficients in Eq. (18) and Eq. (19).

We assume that the expansion coefficients of the **reflected** scattered field are able to be linked with the ones of the **direct** scattered field,

$$\begin{pmatrix}\mathbf{a}^r\\\mathbf{b}^r\end{pmatrix} = \mathbf{R}\cdot\begin{pmatrix}\mathbf{a}^s\\\mathbf{b}^s\end{pmatrix}. \quad (21)$$

The steps towards the **R**-matrix are explained later in Section III.A. By combining Eq. (20), Eq. (21) with Eq. (14), we reach the main equation behind the proposed tool,

$$\begin{pmatrix}\mathbf{a}^s\\\mathbf{b}^s\end{pmatrix} = (\mathbf{1}-\mathbf{T}\cdot\mathbf{R})^{-1}\cdot\mathbf{T}\cdot\begin{pmatrix}\mathbf{a}^i\\\mathbf{b}^i\end{pmatrix}. \quad (22)$$

In Eq. (22), it is seen that, given that the external excitations are assumed to be known, the expansion coefficients of the **direct** scattered field are the main **unknown** of the equation. Once solved, they serve as the starting point to recover the total field everywhere in space (see Section III.A).

Although Eq. (22) looks like the one for the local response case (e.g., Eq. (2.203) on Page 168 in [53]), the use of the two non-classical material response models, i.e., the NLHDM or the SRM, has a significant impact on the evaluation of the **T**- and the **R**-matrix. For the NLHDM, the longitudinal waves (being curl-free) must be included in addition to the transverse waves (being divergence-free). Together with the associated ABC(s), this definitely affects the evaluation of the T-matrix for the NS and the reflection and transmission of PWs through the layers [34], [41]. Likewise, for the SRM, a systematic adaptation must be done according to the quantum corrected BCs in Eq. (12) and Eq. (13) for both the **T**- and the **R**-matrix. The needed adaptations are deliberated in Section III.

### III. Implementation

In this section, bearing the NLHDM and the SRM in mind, we focus on (1) the derivation of the **R** matrix and (2) with the emphasis on the SRM, an S-matrix formalism which deals with the reflection and transmission of waves through multiple spherical and planar interfaces.

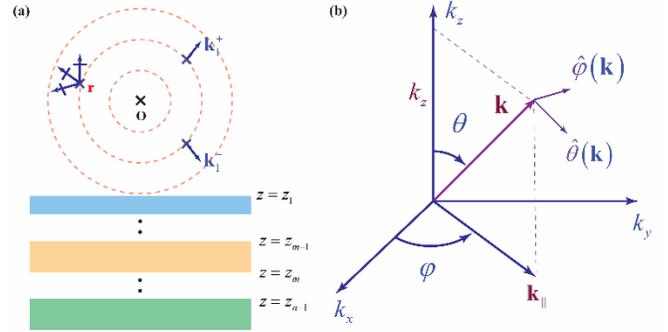

Fig. 2. Illustration of (a) a plane wave expansion of spherical waves and (b) the **k** space. In (a), the dashed circles mark the spherical waves; the center of the spherical waves is set as the origin of the coordinate system; and the spherical waves interact with the *N*-layer substrate. At a spatial point **r**, the spherical waves can be expanded in terms of a spectrum of plane waves as in Eq. (23). In (a), two plane wave components, i.e., an up-propagating, i.e., $\mathbf{k}_1^+$, and a down-propagating, i.e., $\mathbf{k}_1^-$, plane wave, are highlighted. In (b), the **k** space where a wave-vector lives is demonstrated.

#### A. The **R**-Matrix

##### A.1. Expansion in terms of Plane Waves

The first step begins with the expansion of *radiating* SWs which are the bases of the **direct** scattered field, as in Eq. (16), in terms of PWs [62],

$$\begin{pmatrix}\mathbf{M}_{nm}^s(k_1,\mathbf{r})\\\mathbf{N}_{nm}^s(k_1,\mathbf{r})\end{pmatrix} = \frac{1}{2\pi i^n}\cdot$$
$$\int_0^{2\pi}\int_0^{+\infty}\left[\mathbf{a}(\mathbf{k}_1^\pm)\cdot\hat{\theta}(\mathbf{k}_1^\pm) + \mathbf{b}(\mathbf{k}_1^\pm)\cdot\hat{\varphi}(\mathbf{k}_1^\pm)\right]e^{i\mathbf{k}_1^\pm\cdot\mathbf{r}}\frac{k_\rho dk_\rho d\varphi}{k_{1z}k_1}. \quad (23)$$

In Eq. (23), $\mathbf{k}_1^\pm$ is a wave vector in the top layer (see Fig. 2(a) and (b)),

$$\mathbf{k}_1^\pm = \mathbf{k}_\parallel \pm k_{1z}\mathbf{z}, \quad \mathbf{k}_\parallel = k_x\mathbf{x} + k_y\mathbf{y}, \quad (24)$$

$$k_\rho = \sqrt{k_x^2 + k_y^2}, \quad k_{1z} = \sqrt{k_1^2 - k_\rho^2}. \quad (25)$$

Since in Eq. (23) the integration with respect to $k_\rho$ extends to infinity, both the propagating (where $k_{1z}$ is a real number) and the evanescent (where $k_{1z}$ is an imaginary number) spectrum are considered. To ensure the Sommerfeld radiation condition (e.g., chapter 2 in [61]), the square root takes the branch in which the imaginary part of $k_{1z}$ is always positive. Also, the $\pm$ sign corresponds to a wave traveling along the positive or the negative $z$ direction. The $\hat{\theta}$ and $\hat{\varphi}$ are unit vectors (see Fig.

2(b)) transverse to **k** and correspond to the TM (*p*-polarized) and the TE (*s*-polarized) waves, respectively. The amplitudes of the TM and TE waves are summarized in two 2 by 1 column vectors,

$$\mathbf{a}(\mathbf{k}_1^\pm) = \begin{pmatrix} \tilde{\pi}_n^m[\theta(\mathbf{k}_1^\pm)] \\ i\tilde{\tau}_n^m[\theta(\mathbf{k}_1^\pm)] \end{pmatrix} e^{im\varphi(\mathbf{k}_1^\pm)}, \quad (26)$$

$$\mathbf{b}(\mathbf{k}_1^\pm) = \begin{pmatrix} -\tilde{\tau}_n^m[\theta(\mathbf{k}_1^\pm)] \\ i\tilde{\pi}_n^m[\theta(\mathbf{k}_1^\pm)] \end{pmatrix} e^{im\varphi(\mathbf{k}_1^\pm)}. \quad (27)$$

In Eq. (26) and Eq. (27), $\theta$ and $\varphi$ (see Fig. 2(b)) are the elevation and the azimuthal angles in the **k** space. The $\tilde{\pi}$ and $\tilde{\tau}$ functions are defined as,

$$\tilde{\pi}_n^m(\theta) = im \cdot \frac{N_{nm} P_n^m(\cos\theta)}{\sin\theta}, \quad (28)$$

$$\tilde{\tau}_n^m(\theta) = N_{nm} \frac{d}{d\theta} P_n^m(\cos\theta), \quad (29)$$

$$\tilde{P}_n^m(\theta) = N_{nm} P_n^m(\cos\theta). \quad (30)$$

Here, $P_n^m$ is the associated Legendre polynomial [63] and $N_{nm}$ are normalized constants,

$$N_{nm} = \sqrt{\frac{(n-m)!}{(n+m)!} \frac{2n+1}{4\pi}}. \quad (31)$$

*A.2. Reflection and Transmission of Plane Waves*

In the second step, we trace the reflection and transmission of each PW in the expansion of Eq. (23). In the top layer (assumed to be local and isotropic), the reflected waves are,

$$r_p(k_\rho) \cdot \mathbf{a}(\mathbf{k}_1^-) \cdot \hat{\theta}(\mathbf{k}_1^+) e^{i\mathbf{k}_\parallel \cdot \mathbf{r}_\parallel} e^{+ik_{1z}(z-2z_1)}, \quad (32)$$

$$r_s(k_\rho) \cdot \mathbf{b}(\mathbf{k}_1^-) \cdot \hat{\varphi}(\mathbf{k}_1^+) e^{i\mathbf{k}_\parallel \cdot \mathbf{r}_\parallel} e^{+ik_{1z}(z-2z_1)}. \quad (33)$$

Here, $\mathbf{r}_\parallel = (x, y)$ and $z_1$ is the position of the first interface (see Fig. 2(a)). In the bottom layer, the transmitted waves are,

$$t_p(k_\rho) \cdot \mathbf{a}(\mathbf{k}_1^-) e^{-ik_{1z}z_1} \cdot \hat{\theta}(\mathbf{k}_N^-) e^{i\mathbf{k}_\parallel \cdot \mathbf{r}_\parallel} e^{-ip_{nz}(z-z_{n-1})}, \quad (34)$$

$$t_s(k_\rho) \cdot \mathbf{b}(\mathbf{k}_1^-) e^{-ik_{1z}z_1} \cdot \hat{\varphi}(\mathbf{k}_N^-) e^{i\mathbf{k}_\parallel \cdot \mathbf{r}_\parallel} e^{-ik_{nz}(z-z_{n-1})}. \quad (35)$$

Here, $z_{n-1}$ is the position of the last interface (see Fig. 2(a)). In Eq. (35), it is noted that we distinguish $p_{nz}$ and $k_{nz}$ as the $z$ component of the wave vector. This is for the case where the bottom layer is filled by a uniaxial medium. In detail, $p_{nz}$ and $k_{nz}$ are,

$$p_{nz} = \sqrt{k_n^2 - (\varepsilon_{n,t}/\varepsilon_{n,z})k_\rho^2}, \; k_{nz} = \sqrt{k_n^2 - k_\rho^2}, \; k_n^2 = \varepsilon_{n,t} k_0^2. \quad (36)$$

In Eq. (36), $\varepsilon_{n,t}$ and $\varepsilon_{n,z}$ are the in-plane and out-of-plane permittivity of the $n^{th}$ layer (the bottom layer). In a mid-layer, i.e., the $m^{th}$ layer, the scattered PWs are,

$$c_p(+p_{mz}) \cdot \mathbf{a}(\mathbf{k}_1^-) e^{-ik_{1z}z_1} \cdot \hat{\theta}(\mathbf{k}_m^+) e^{i\mathbf{k}_\parallel \cdot \mathbf{r}_\parallel} e^{+ip_{mz}(z-z_m)} + $$
$$c_p(-p_{mz}) \cdot \mathbf{a}(\mathbf{k}_1^-) e^{-ik_{1z}z_1} \cdot \hat{\theta}(\mathbf{k}_m^-) e^{i\mathbf{k}_\parallel \cdot \mathbf{r}_\parallel} e^{-ip_{mz}(z-z_{m-1})}, \quad (37)$$

$$c_s(+k_{mz}) \cdot \mathbf{a}(\mathbf{k}_1^-) e^{-ik_{1z}z_1} \cdot \hat{\varphi}(\mathbf{k}_m^+) e^{i\mathbf{k}_\parallel \cdot \mathbf{r}_\parallel} e^{+ik_{mz}(z-z_m)} + $$
$$c_s(-k_{mz}) \cdot \mathbf{a}(\mathbf{k}_1^-) e^{-ik_{1z}z_1} \cdot \hat{\varphi}(\mathbf{k}_m^-) e^{i\mathbf{k}_\parallel \cdot \mathbf{r}_\parallel} e^{-ik_{mz}(z-z_{m-1})}. \quad (38)$$

Here, $z_{m-1}$ and $z_m$ are the positions of the upper and the lower interfaces of the $m^{th}$ layer (see Fig. 2(b)). Again, $p_{mz}$ and $k_{mz}$ are distinguished, for the case the medium filling the $m^{th}$ layer is uniaxial.

Also, if the $m^{th}$ layer is modelled by the NLHDM, an additional longitudinal wave should be taken into account,

$$c_l(+\kappa_{mz}) \cdot \mathbf{a}(\mathbf{k}_1^-) e^{-ik_{1z}z_1} \cdot \hat{\kappa}(\boldsymbol{\kappa}_m^+) e^{i\mathbf{k}_\parallel \cdot \mathbf{r}_\parallel} e^{+i\kappa_{mz}(z-z_m)} + $$
$$c_l(-\kappa_{mz}) \cdot \mathbf{a}(\mathbf{k}_1^-) e^{-ik_{1z}z_1} \cdot \hat{\kappa}(\boldsymbol{\kappa}_m^-) e^{i\mathbf{k}_\parallel \cdot \mathbf{r}_\parallel} e^{-i\kappa_{mz}(z-z_{m-1})}. \quad (39)$$

In Eq. (39), $\hat{\kappa}$ marks a unit vector along the wave vector $\boldsymbol{\kappa}_m$,

$$\boldsymbol{\kappa}_m^\pm = \mathbf{k}_\parallel \pm \kappa_{1z}\mathbf{z}, \; k_{mz} = \sqrt{l_m^2 - k_\rho^2}. \quad (40)$$

Here, $l$ is known as the longitudinal wave number,

$$l = \frac{1}{\beta}\left[\omega(\omega + i\gamma) - \frac{\omega_p^2}{\varepsilon_{bd}}\right]. \quad (41)$$

We note that, when the $m^{th}$ layer is the bottom layer, the first term, i.e., for a PW propagating along the positive $z$ direction, in Eq. (39) should be removed. Also, the evaluation for the coefficients $r_s$, $r_p$, $t_s$, $t_p$, $c_p$, $c_s$ and $c_l$ will be discussed in Section III.B.

*A.3. Integration*

The third step collects the effects of all reflected, scattered and transmitted PWs. Based on Eq. (32), Eq. (33) and Eq. (23), the **reflected** SWs in the top layer are,

$$\begin{pmatrix} \mathbf{M}_{nm}^r(\mathbf{r}) \\ \mathbf{N}_{nm}^r(\mathbf{r}) \end{pmatrix} = \frac{1}{2\pi i^n} \cdot $$
$$\left\{ \int_0^{2\pi} \int_0^{+\infty} \left[ r_p(k_\rho) \cdot \mathbf{a}(\mathbf{k}_1^-) \cdot \hat{\theta}(\mathbf{k}_1^+) e^{i\mathbf{k}_\parallel \cdot \mathbf{r}_\parallel} e^{+ik_{1z}(z-2z_1)} + \right. \right. \quad (42)$$
$$\left. \left. r_s(k_\rho) \cdot \mathbf{b}(\mathbf{k}_1^-) \cdot \hat{\varphi}(\mathbf{k}_1^+) e^{i\mathbf{k}_\parallel \cdot \mathbf{r}_\parallel} e^{+ik_{1z}(z-2z_1)} \right] \frac{k_\rho dk_\rho d\varphi}{k_{1z}k_1} \right\}.$$

Based on Eq. (34), Eq. (35) and Eq. (23), the **transmitted** SWs in the bottom layer are,

$$\begin{pmatrix} \mathbf{M}_{nm}^t(\mathbf{r}) \\ \mathbf{N}_{nm}^t(\mathbf{r}) \end{pmatrix} = \frac{1}{2\pi i^n} \cdot $$
$$\left\{ \int_0^{2\pi} \int_0^{+\infty} \left[ t_p(k_\rho) \cdot \mathbf{a}(\mathbf{k}_1^-) \cdot \hat{\theta}(\mathbf{k}_N^-) e^{i\mathbf{k}_\parallel \cdot \mathbf{r}_\parallel} e^{-ik_{1z}z_1 - ip_{nz}(z-z_{N-1})} + \right. \right. \quad (43)$$
$$\left. \left. t_s(k_\rho) \cdot \mathbf{b}(\mathbf{k}_1^-) \cdot \hat{\varphi}(\mathbf{k}_N^-) e^{i\mathbf{k}_\parallel \cdot \mathbf{r}_\parallel} e^{-ik_{1z}z_1 - ik_{nz}(z-z_{N-1})} \right] \frac{k_\rho dk_\rho d\varphi}{k_{1z}k_1} \right\}.$$

Based on Eq. (37), Eq. (38) and Eq. (23), the **scattered** SWs in a mid-layer, e.g., in the $m^{th}$ layer, are,

$$\begin{pmatrix} \mathbf{M}_{nm}^{sca}(\mathbf{r}) \\ \mathbf{N}_{nm}^{sca}(\mathbf{r}) \end{pmatrix} = \frac{1}{2\pi i^n} \cdot \sum_\pm $$
$$\left\{ \int_0^{2\pi} \int_0^{+\infty} \left[ c_p(\pm k_{mz}) \cdot \mathbf{a}(\mathbf{k}_1^-) \cdot \hat{\theta}(\mathbf{k}_m^\pm) e^{i\mathbf{k}_\parallel \cdot \mathbf{r}_\parallel} e^{-ik_{1z}z_1 \pm ip_{mz}(z-z_m)} + \right. \right. \quad (44)$$
$$\left. \left. c_s(\pm k_{mz}) \cdot \mathbf{a}(\mathbf{k}_1^-) \cdot \hat{\varphi}(\mathbf{k}_m^\pm) e^{i\mathbf{k}_\parallel \cdot \mathbf{r}_\parallel} e^{-ik_{1z}z_1 \pm ik_{mz}(z-z_m)} \right] \frac{k_\rho dk_\rho d\varphi}{k_{1z}k_1} \right\}.$$

In Eq. (44), the summation is to sum up the PWs propagating along the positive and the negative $z$ direction in the $m^{th}$ layer.

When the $m^{th}$ layer is modelled by the NLHDM, besides the transverse SWs in Eq. (44), based on Eqs. (39) and (23), a longitudinal scattered spherical wave should be considered,

$$\mathbf{L}_{nm}^{(R)}(\mathbf{r}) = \frac{1}{2\pi i^n} \cdot \sum_{\pm} \left[ \int_0^{2\pi} \int_0^{+\infty} c_l(\pm\kappa_{mz}) \cdot \mathbf{a}(\mathbf{k}_1^-) e^{-ik_{1z}z_1} \cdot \hat{\kappa}(\kappa_m^\pm) e^{i\mathbf{k}_\parallel \cdot \mathbf{r}_\parallel} e^{\pm i\kappa_{mz}(z-z_m)} \frac{k_\rho dk_\rho d\varphi}{k_{1z}k_1} \right]. \quad (45)$$

In Eq. (45), the summation is the same as Eq. (44).

As a remark, since the origin is set at the center of the NS, $z_1$ is a negative real number. The phase term $e^{-ik_{1z}z_1}$ in Eq. (42) - Eq. (45) decays exponentially as $k_\rho$ approaches infinity. Thus, the numerical convergence of the integrals in these equations is always guaranteed.

Once the expansion coefficients in Eq. (22) are solved, the reflected, the transmitted and the scattered SWs in Eq. (42) – Eq. (45) serve as the bases to reconstruct the scattered fields in all layers. As an example, the **reflected** scattered field in the top layer can be written as,

$$\mathbf{E}^r(\mathbf{r}) = \sum_{nm} \left[ \mathbf{M}_{nm}^r(\mathbf{r}) \cdot a_{nm}^s + \mathbf{N}_{nm}^r(\mathbf{r}) \cdot b_{nm}^s \right]. \quad (46)$$

To be complete, in our implementation, the integration with respect to $\varphi$ is done analytically. This is explained in detail in Appendix I.

*A.4. The **R** Matrix*

To obtain the **R** matrix, the **reflected** scattered field in the top layer (see Eq. (46)) must be expanded in terms of *standing* SWs, which is seen by comparing Eq. (19) with Eq. (46). The desired expansion is done by noting that the TM and the TE waves in the integral kernel of Eq. (42) can be expanded in terms of *standing* SWs (see e.g., [53] and in our previous work [41]),

$$\begin{pmatrix} \hat{\theta}(\mathbf{k}_1^+) \\ \hat{\varphi}(\mathbf{k}_1^+) \end{pmatrix} e^{i\mathbf{k}_1^+ \cdot \mathbf{r}} = \sum_{n'm'} \frac{4\pi i^{n'}}{n'(n'+1)} \cdot \begin{pmatrix} \tilde{\pi}_{n'}^{m'}(\theta(\mathbf{k}_1^+)) & i\tilde{\tau}_{n'}^{m'}(\theta(\mathbf{k}_1^+)) \\ \tilde{\tau}_{n'}^{m'}(\theta(\mathbf{k}_1^+)) & -i\tilde{\pi}_{n'}^{m'}(\theta(\mathbf{k}_1^+)) \end{pmatrix} e^{-im'\varphi(\mathbf{k}_1^+)} \cdot \begin{pmatrix} \mathbf{M}_{n'm'}^e(k_1,\mathbf{r}) \\ \mathbf{N}_{n'm'}^e(k_1,\mathbf{r}) \end{pmatrix}. \quad (47)$$

By substituting Eq. (47) in Eq. (42), the **R** matrix is reached,

$$\begin{pmatrix} \mathbf{M}_{nm}^r(k_1,\mathbf{r}) \\ \mathbf{N}_{nm}^r(k_1,\mathbf{r}) \end{pmatrix} = \sum_{n'm'} \mathbf{R}_{nm,n'm'} \cdot \begin{pmatrix} \mathbf{M}_{n'm'}^e(k_1,\mathbf{r}) \\ \mathbf{N}_{n'm'}^e(k_1,\mathbf{r}) \end{pmatrix}. \quad (48)$$

In Eq. (48), $\mathbf{R}_{nm,n'm'}$ is

$$\mathbf{R}_{nm,n'm'} = \frac{4\pi i^{n'-n}\delta_{mm'}}{n'(n'+1)} \cdot$$

$$\int_0^{+\infty} \begin{pmatrix} \tilde{\pi}_n^m[\theta(\mathbf{k}_1^-)] & -\tilde{\tau}_n^m[\theta(\mathbf{k}_1^-)] \\ i\tilde{\tau}_n^m[\theta(\mathbf{k}_1^-)] & i\tilde{\pi}_n^m[\theta(\mathbf{k}_1^-)] \end{pmatrix} \cdot \begin{pmatrix} r_p(k_\rho) & 0 \\ 0 & r_s(k_\rho) \end{pmatrix} \cdot \begin{pmatrix} -\tilde{\pi}_{n'}^{m'}(\theta(\mathbf{k}_1^+)) & -i\tilde{\tau}_{n'}^{m'}(\theta(\mathbf{k}_1^+)) \\ -\tilde{\tau}_{n'}^{m'}(\theta(\mathbf{k}_1^+)) & +i\tilde{\pi}_{n'}^{m'}(\theta(\mathbf{k}_1^+)) \end{pmatrix} e^{-2ik_{1z}z_1} \frac{k_\rho dk_\rho}{k_{1z}k_1} \quad (49)$$

As a comment, the convergence of the semi-infinite integral in Eq. (49) is guaranteed, by noting the phase term $e^{-2ik_{1z}z_1}$ in Eq. (49) and the fact that: (I) the branch of the square root in Eq. (25) has been picked in such a way that the imaginary part of $k_{1z}$ is always a positive real number; and (II) $z_1$ is a real negative number.

*B. Reflection and Transmission of Waves through Spherical and Planar Layers*

By using S-matrices, a unified framework can be developed for both the NS, i.e., a spherically layered system, and the "mirror", i.e., a planarly layered system. The key to the framework is to find the S-matrix for each spherical or planar interface. Then, the T-matrix for the NS in Eq. (22), and the amplitudes of the reflected, the transmitted and the scattered waves in spherical and planar layers (used in Eqs. (32) – (39)) can be retrieved by concatenating the S-matrices. The concatenation is known as the Redheffer star operation [64] and is well-archived in literature (e.g., [36], [38], [41]) and thus will not be repeated here. Also, it is noted that the case in which the layers are modelled by the NLHDM is well-covered in our previous work [41]. Thus, in the following, we will focus on the S-matrix for a single interface within the SRM.

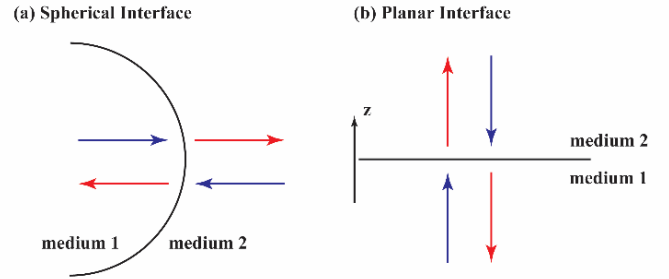

Fig. 3. Illustration of (a) a spherical and (b) a planar interface. The interfaces separate an inner region filled by medium 1 from an outer medium filled by medium 2. In both figures, the outgoing waves are marked by the red color, while the incoming ones are marked by the blue color. In (b), the positive $z$ direction is marked by an arrow.

According to the QC – BCs in Eq. (12) and Eq. (13) for the SRM, a matrix equation can be listed for an interface,

$$\mathbf{Q}_1^+ \cdot c_1^+ + \mathbf{Q}_1^- \cdot c_1^- = \mathbf{Q}_2^+ \cdot c_2^+ + \mathbf{Q}_2^- \cdot c_2^-. \quad (50)$$

In Eq. (50), **Q**'s and $c$'s are wave matrices and wave amplitudes. The subscripts "1" and "2" mark the inner region and the outer region of the interface and the superscripts "+" and "-" do the outgoing and the incoming types of waves (see the red and blue arrows in Fig. 3). By reshuffling the terms on both sides of Eq. (50), the S-matrix can be obtained as,

$$\mathbf{S} = \begin{bmatrix} +\mathbf{Q}_1^- & -\mathbf{Q}_2^+ \end{bmatrix}^{-1} \cdot \begin{bmatrix} -\mathbf{Q}_1^+ & +\mathbf{Q}_2^- \end{bmatrix}. \quad (51)$$

The S-matrix links the amplitudes of the incoming waves with the ones of the outgoing waves,

$$\begin{bmatrix} c_1^- \\ c_2^+ \end{bmatrix} = \mathbf{S} \cdot \begin{bmatrix} c_1^+ \\ c_2^- \end{bmatrix}. \quad (52)$$

The mathematical forms of **Q** and c in Eqs. (50) – (52) depend on whether a spherical or a planar interface is considered.

For the former, **Q** takes a generic form (see the derivations in Appendix II), for the TE system, corresponding to the **M** function in Eq. (15) and Eq. (16),

$$\mathbf{Q} = \begin{pmatrix} z_n(kr) \\ \frac{1}{iZ} \cdot \frac{1}{kr} \frac{\partial(rz_n(kr))}{\partial r} + i\omega d_\parallel \cdot \varepsilon \cdot z_n(kr) \end{pmatrix}, \quad (53)$$

for the TM system, which corresponds to the **N** function in Eq. (15) and Eq. (16),

$$\mathbf{Q} = \begin{pmatrix} \frac{1}{kr}\frac{\partial(rz_n(kr))}{\partial r} + d_\perp \cdot n(n+1) \cdot \frac{z_n(kr)}{kr^2} \\ \frac{1}{iZ} \cdot z_n(kr) + i\omega d_\parallel \cdot \varepsilon \cdot - \frac{1}{kr}\frac{\partial(rz_n(kr))}{\partial r} \end{pmatrix}. \quad (54)$$

The expansion coefficient $c$'s corresponding to the TE and the TM systems are $a_{nm}$ and $b_{nm}$ as in Eqs. (15) and (16). In Eqs. (53) and (54), $r$ is the radius of the interface. $k$, $Z$ and $\varepsilon$ are the wavenumber, the wave impedance, and the permittivity of the material. Their values depend on whether the inner region (the subscript being "1") or the outer region (the subscript being "2") is regarded. Lastly, $z_n(kr)$ can be the spherical Bessel (for the superscript being "-") or the spherical Hankel function (for the superscript being "+").

For the latter, **Q** takes a generic form (see the derivations in Appendix II), for the TE system (i.e., the TE wave and also see the $\hat{\varphi}$ vector in Fig. 2(b)),

$$\mathbf{Q} = \begin{pmatrix} 1 \\ -\frac{k_z}{\omega\mu_0} - i\omega d_\parallel \cdot \varepsilon_0\varepsilon_t \end{pmatrix} e^{i\mathbf{k}_\parallel \cdot \mathbf{r}_\parallel + ik_z(z-z_0)}, \quad (55)$$

for the TM system (i.e., the TM wave and also see the $\hat{\theta}$ vector in Fig. 2(b)),

$$\mathbf{Q} = \begin{pmatrix} \frac{q_z}{q} - \frac{\varepsilon_t}{\varepsilon_z}\frac{ik_\rho^2 d_\perp}{q} \\ \frac{\omega\varepsilon_0\varepsilon_t}{q}(1 + id_\parallel q_z) \end{pmatrix} e^{i\mathbf{k}_\parallel \cdot \mathbf{r}_\parallel + iq_z(z-z_0)}. \quad (56)$$

The $c$'s are $a(k_z)$ and $b(q_z)$ for the TE and the TM systems, respectively (see Appendix II for more details). We note that Eqs. (55) and (56) are generalized to include uniaxial media (e.g., graphene layers). This contrasts to the works [26], [27] where isotropic media are considered and is reflected in Eqs. (55) and (56) by the in-plane and out-of-plane permittivity, i.e., $\varepsilon_t$ and $\varepsilon_z$, in the definition of $k$, i.e., $k^2 = \varepsilon_t k_0^2$, and lastly in the use $k_z$ and $q_z$ for the TE and the TM waves.

In Eq. (55) and Eq. (56), the interface is assumed to be at $z = z_0$. Like the spherical case, the value of the wavenumber $k$ and the permittivity $\varepsilon_t$ and $\varepsilon_z$ depends on whether the inner (the subscript being "1") or the outer (the subscript being "2") is considered. Also, $k_z$ and $q_z$ are place holders for $\pm k_z$ and $\pm q_z$ highlighting the upgoing (for the superscript being "+", i.e., propagation along the positive $z$ direction, see Fig. 3(b) for the coordinate system) and the down-going (for the subscript being "-", i.e., propagation along the negative $z$ direction, see Fig. 3(b) for the coordinate system) waves.

IV. RESULTS

In this section, we verify the implementation by comparing the results from the proposed tool with the ones from an in-house developed BEM solver [47], [49], [52]. Then, two further examples are presented to demonstrate the physical impact of the NLHDM and the SRM on the optical response (both near field and far field) of the NSoM structure. The simulations are run on a workstation with a 16-core CPU (Ryzen 7950X) and 128 GB RAM.

*A. A Quantitative Check*

In the example, we consider an NS with a radius of 20 nm. The NS is made of Gold (Au) and is positioned on top of an Au mirror (see Fig. 4). The NS and the mirror are separated by a 1nm gap. The excitation is a TM-polarized oblique incident plane wave with an incidence angle of 60° (see the inset of Fig. 4). For Au, the permittivity is from tabulated data [65], while the Fermi velocity $v_f$ is $1.40 \times 10^6$ m/s. The 1nm gap is assumed to be filled in by a material with refractive index 1.5. The far field radiated by the NSoM is calculated on a hemi-sphere with a radius of 1 m in the top layer (see the black dash line in the inset of Fig. 4). The hemi-sphere is discretized by 1569 triangles and the electric field, the magnetic field, and the Poynting vector are evaluated at the centroids of the triangles, based on which the scattered power collected by the hemisphere is calculated. This scheme of evaluating far field properties will be used for the rest of this work. Finally, as a remark, for the evaluation of the far field, the reflected SWs in Eq. (42) are needed. There, numerical integrations are avoided by using the stationary phase method [61].

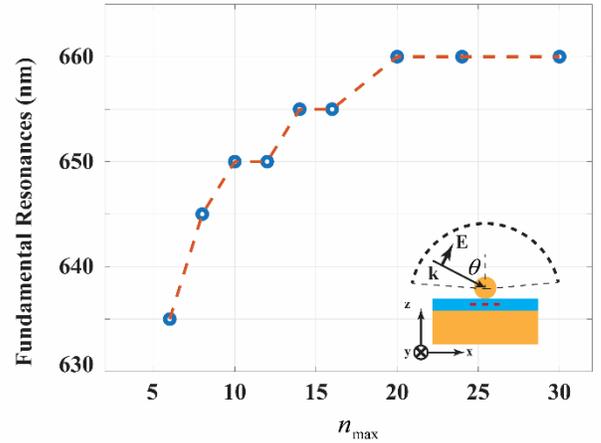

Fig. 4. Convergence test for the T-matrix solver. The spectral position of the fundamental resonance in the scattered power spectra of the NSoM structure is plotted against the maximum number of SWs used in the simulation. In the inset of the figure, the simulated structure is shown and especially the cut in Fig. 5(b) – (e) where the electric field is evaluated is marked by the red dashed line. The wave vector **k** of the incident plane wave forms an angle $\theta = 60°$ with respect to the vertical direction and the incident wave is TM-polarized. The hemi-sphere where the far field is collected is denoted by the black dash line.

For the proposed tool, different numbers of SWs are used to test the convergence of the results. In Fig. 4, we plot the spectral position of the fundamental resonance (extracted from the scattered power spectra from the NSoM structure) when different numbers of SWs, i.e., $n_{\max}$ in Fig. 4, are used. It is seen that, for $n_{max} \geq 20$ (we test up to $n_{max} = 30$), the spectral position converges at 660 nm.

Further, we compare the results from the proposed tool ($n_{max} = 20$) with the ones from the BEM solver. The BEM solver is a dedicated solver for the NPoM structure and can properly account for the NLHDM [52]. The BEM solver focuses on the boundary of the NS and discretizes the boundary into small triangular patches. In the example, we discretize the boundary by 878, 1246 and 1678 triangles. For both the T-matrix and the BEM simulations, a wavelength range spanning from 570 nm to 800 nm (where the main resonance is located) is considered. The range has 24 sampling points in-between. In both simulations, the center of the NS is set as the origin of the coordinate system.

It can be seen from Fig. 5(a) that by using denser meshes, the scattered power calculated by the BEM solver approaches the one by the proposed tool (see the circled lines and the blue solid line in Fig. 5(a)). Further, we plot the near fields at a cut right in the middle of the gap. The cut is along the $xy$ plane (see the coordinate system and the position of the cut in the inset of Fig. 4) is with a size of 40 nm by 40 nm. 41 sampling points are taken along each direction. It can be seen from Fig. 5(b) to (e) that, again with denser meshes, the BEM results approach the ones from the proposed tool. To be concrete, we evaluate the average and the maximum relative errors,

$$\frac{1}{N}\sum_{i=1}^{N}\frac{\left\| \mathbf{E}^{BEM}(\mathbf{r}_i)\right| - \left| \mathbf{E}^{T}(\mathbf{r}_i) \right\|}{\left| \mathbf{E}^{T}(\mathbf{r}_i) \right|}, \quad (57)$$

$$\max\left\{ \frac{\left\| \mathbf{E}^{BEM}(\mathbf{r}_i)\right| - \left| \mathbf{E}^{T}(\mathbf{r}_i) \right\|}{\left| \mathbf{E}^{T}(\mathbf{r}_i) \right|} \right\}. \quad (58)$$

In Eq. (57) and Eq. (58), the superscripts "BEM" and "T" mark the results from the BEM solver and the proposed tool, respectively. The subscript "$i$" refers to the $i^{th}$ sampling point in the cut. $N$ is the total number of sampling points on the cut. |**E**| takes the magnitude of the electric field and "max" picks out the maximum value. For the three meshes, the average relative errors are 0.0603, 0.0307 and 0.0109, while the maximum relative errors are 0.2678, 0.1903 and 0.1301.

*B. Physical Checks*

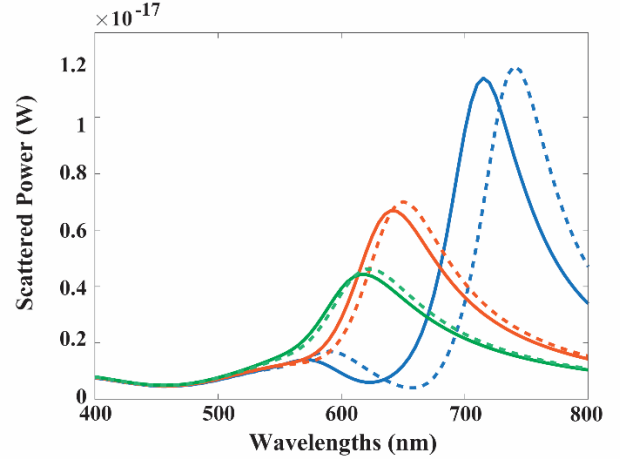

Fig. 6. The power scattered from the NSoM structure with different gap sizes. The G = 1 nm case, the G = 3 nm case and the G = 5 nm case are marked by the blue, orange and green colors, respectively. In the plot, the dashed lines and the solid lines correspond to the local and the nonlocal case, respectively.

As a physical check, we compare the impact of the LRM with the one of the NLHDM on the optical response of the NSoM. To illustrate, we consider an Au NS with a radius of 30 nm on Au mirror with various gap sizes, i.e., 1 nm, 3 nm, and 5 nm. The gap is still filled by a medium with a refractive index of 1.5. The NSoM structure is excited by an oblique incident plane wave (the incident direction forming a 60° angle with respect to the vertical axis). The whole setup resembles the one in the inset of Fig. 4. The considered wavelength is from 400 nm to 800 nm with 81 sampling points in between. Here, a sufficient number of SWs ($n_{max} = 20$ for all gap sizes) is used.

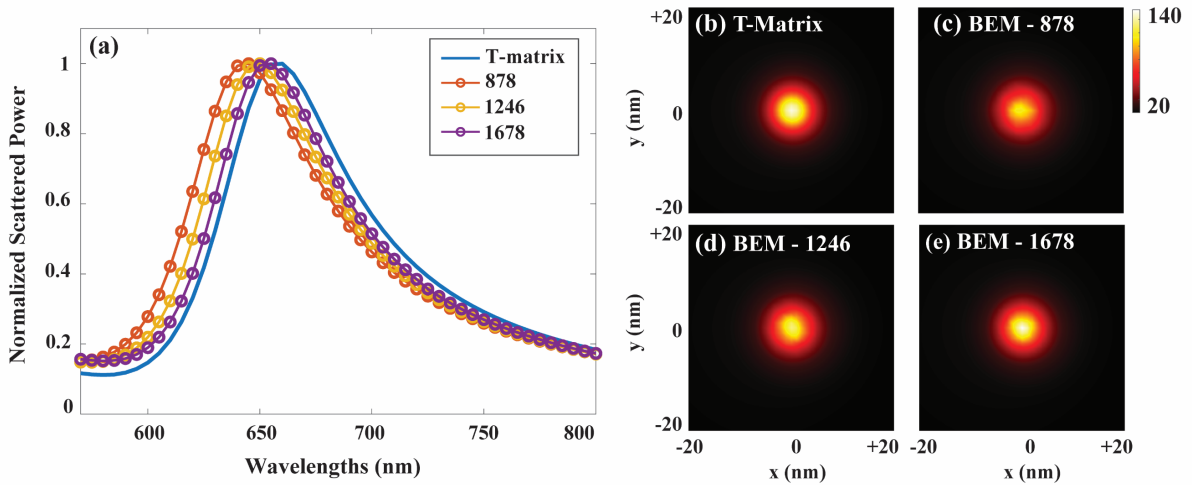

Fig. 5. Comparison between the results from the proposed tool and the BEM solver. The result from the T-matrix method is compared with the ones from the BEM where 878, 1246 and 1678 triangles are used. (a) plots the normalized scattered power, i.e., the scattered power from the T-matrix method and the BEM solver is normalized with respect to the maximum of the scattered power. In (b) – (e), the magnitude of the scattered electric field on a 40 nm by 40 nm cut (with 41 sampling points in each direction and see the position of the cut in the inset of Fig. 4) is plotted. Here, the magnitude of the scattered electric field is normalized with respect to the one of the incident electric field. The color is coded from black to red to mark the amplitude of the electric field.

It can be seen from Fig. 6 that the main resonance in the scattered power spectrum predicted by the NLHDM is always blue shifted with respect to the one by the LRM. This is physically due to the "spill-in" of charges predicted by the NLHDM model which reduces the electrical size of the NS. Further, the amount of the blue shift reduces as the gap size increases, marking the importance of non-classical effects at the deep-nm level.

In the near field regime, similar to the previous example, the electric field is calculated on a cut right in the middle of the gap (similar to the one in the inset of Fig. 4). The cut is along the $xy$ plane (see the coordinate system in the inset of Fig. 4) and spans an area of 60 nm by 60 nm. 61 sampling points are taken along each direction. In general, it is seen from Fig. 7 that, as the gap size increases, the electric field enhancement decreases. This is due to the reduced coupling between the NS and its image. The enhancement predicted by the NLHDM is always weaker than the one by the LRM (see Fig. 7). As discussed in Section II.A, this is a direct result of the collapse of the concept of "surface charge" in classical electrodynamics [58]. Instead, a volume charge distribution near the boundary must be considered. As the gap size increases, the effects arising from the boundary region become less important. Hence, the results from the NLHDM degenerate to the one from LRM at the gap size of 5 nm. The above observations are well in line with previous works [33].

Last but not least, we look at an NS (with a radius of 10 nm) made of a simple metal (a representative metal is Sodium) on top of a mirror made of the same simple metal with a gap size of 0.74 nm. This gap size is very close to the tunneling regime, i.e., 0.5 nm [26]. The bulk of the metal is modelled by the LRM with the plasma and the damping frequency $\omega_p = 5.9$ eV and $\gamma = 0.1$ eV. For the SRM, $d_\perp$, is needed. We adopt a fitting model (see supplementary note 9 in [42]) for the parameter. Since $d_\perp$ is extracted for a vacuum-metal interface, a vacuum gap is considered. Further, the incident light is chosen to be a plane wave with high angle (the incident direction forms an angle of 80° with respect to the vertical axis). Hence, this example can be compared with the one in Fig. 3(d) in [26] where a dimer of Sodium spheres is of the focus. We consider an energy range from 1.5 eV to 6 eV with 251 sampling points in between. A sufficient number of SWs ($n_{max} = 16$) is used. It can be seen from Fig. 8(a) that the resonances in the scattering spectra from the SRM illustrate a systematic red shift against and are broader than the ones from the LRM. The former is due to the "spill-out" of charges which increases the effective electrical length of the NS, while the latter is, similar to the NLHDM case, due to the "diffusive"-like effects at the boundary of the NS. This agrees well with the previous findings in [26]. To be complete, we look at the near field at a cut in the gap (see Fig. 8(b) – (d)). Once again, the field enhancement is reduced.

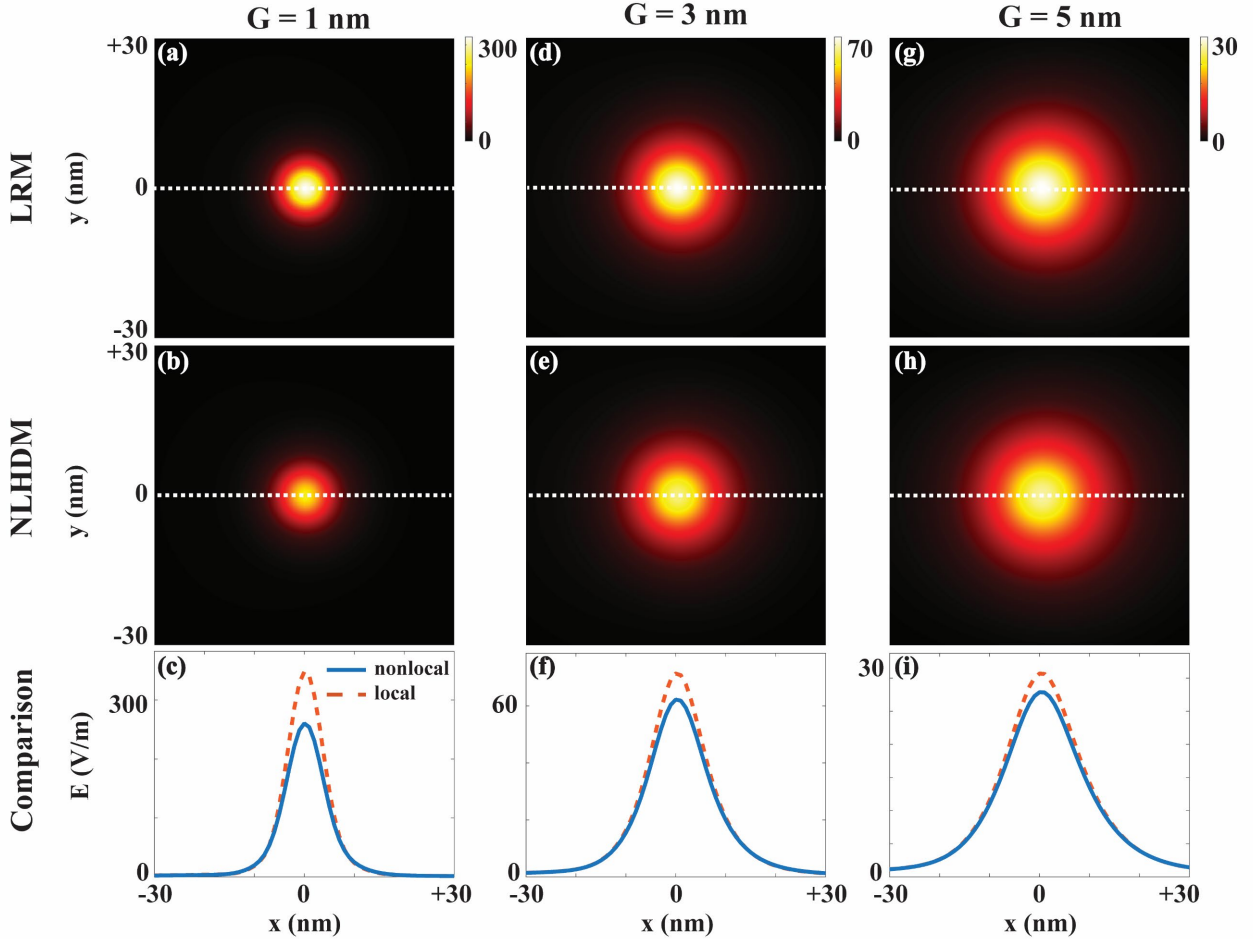

Fig. 7. The magnitude of the scattered electric field on a cut (see the red dashed line in the inset of Fig. 4(a) for the position of the cut) in the mid of the gaps. The cut is with the size of 60 nm by 60 nm with 61 sampling points along each direction. In all plots, the magnitude of the scattered electric field on the cut is plotted, and the magnitude of the scattered electric field is normalized with respect to the one of the incident electric field. The first, the second and the third columns correspond to the cases where G = 1 nm, G = 3 nm, and G = 5 nm, respectively. In (a), (b), (d), (e), (g) and (h), the color is coded from black to yellow to denote the magnitude of the field. In (c), (f) and (i), the magnitudes of the normalized scattered electric field are plotted for the LRM case (the red dashed line) and for the NLHDM case (the blue solid line) along the white dashed lines in (a) and (b), (d) and (e), and (g) and (h), respectively.

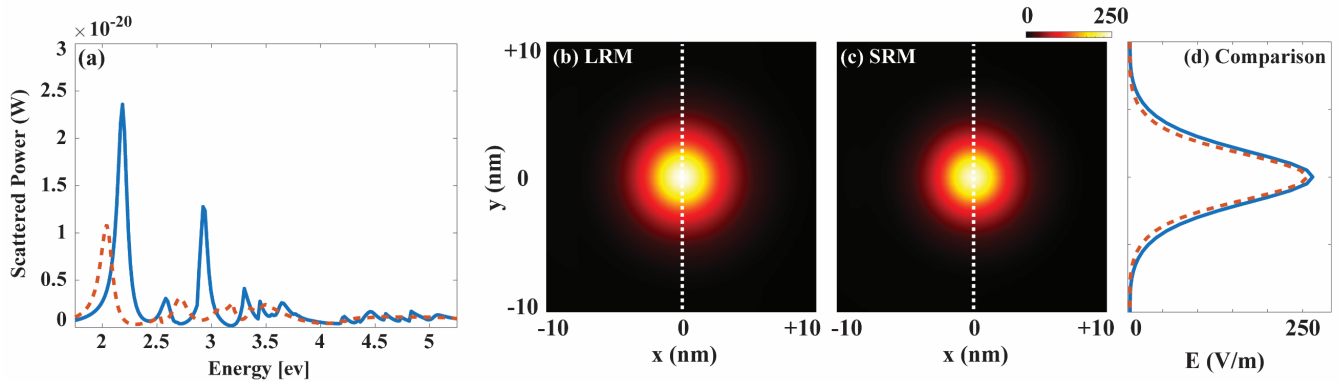

Fig. 8. The power scattered by a Jellium NSoM structure with a gap size of 0.74 nm (a) and the near field enhancement on a cut at the mid of the gap for the LRM (b) and the SRM (c). In (a), the blue solid line represents the scattered power for the LRM case, while the dashed orange line does the scattered power for the SRM case. In (b) and (c), the color is coded from black to yellow to denote the strength of the field. For (b) and (c), the cut spans an area of 20 nm by 20 nm with 41 sampling points being taken in each direction. In (d), a comparison between the LRM case (the blue solid line) and the SRM (the orange dashed line) case is made along the white dashed lines in (b) and (c).

## V. Conclusion

In this work, within the framework of the T-matrix method, we present a dedicated modeling tool for the nanosphere-on-mirror (NSoM) structure where the nonclassical effects in both the sphere and the mirror are accounted for by the nonlocal hydrodynamic model (NLHDM) and the surface response model (SRM). Contrasting with the conventional T-matrix method, we find that two key adaptations must be made: one is for the **T**-matrix, while the other is for the **R**-matrix. The former is done by using the concept of S-matrices, while the latter is resolved by a four-step procedure where conversions between spherical and plane waves are involved. Lastly, by comparing with an in-house developed boundary element method (BEM) solver and with previous physical findings, the proposed tool is quantitatively and qualitatively validated. The proposed modeling tool does not only serve as a reference as the Mie solution for the homogeneous space, but also provides an efficient and effective approach for investigating interesting physics at the deep-nanometric scales and can be become an essential element in the study of mesoscopic electrodynamics.

## Appendix I. Integration with Respect to the $\varphi$ Angle in Eqs. (42) - (45)

In this Appendix, we analytically evaluate the integration with respect to the $\varphi$ angle in Eqs. (42) - (45), so that the original integrals in the equations are reduced from two-dimensional to one-dimensional.

In Eq. (42) – Eq. (45), $\hat{\theta}$, $\hat{\varphi}$ and $\hat{\kappa}$, i.e., the unit vectors of the **k** space (see Fig. 2), are functions of the $\varphi$ angle. Also, the phase terms $e^{i\mathbf{k}_\parallel \cdot \mathbf{r}_\parallel} = e^{ik_\rho \rho \cos(\varphi-\alpha)}$, where $\rho = \sqrt{x^2 + y^2}$ and $\alpha = \operatorname{atan}(y/x)$, and $e^{im\varphi}$ (in the coefficients **a** and **b**, see Eq. (26) and Eq. (27)) are dependent on the $\varphi$ angle. By collecting these terms, we can extract the integrals with respect to the $\varphi$ angle from Eq. (42) – Eq. (45),

$$\int_0^{2\pi} \hat{\theta}(\mathbf{k}) e^{i\mathbf{k}_\parallel \cdot \mathbf{r}_\parallel} e^{im\varphi} d\varphi, \tag{59}$$

$$\int_0^{2\pi} \hat{\varphi}(\mathbf{k}) e^{i\mathbf{k}_\parallel \cdot \mathbf{r}_\parallel} e^{im\varphi} d\varphi, \tag{60}$$

$$\int_0^{2\pi} \hat{k}(\mathbf{k}) e^{i\mathbf{k}_\parallel \cdot \mathbf{r}_\parallel} e^{im\varphi} d\varphi. \tag{61}$$

Then, we realize that the unit vectors can be written as,

$$\hat{\theta}(\mathbf{k}) = \frac{k_z}{k}\cos\varphi \mathbf{x} + \frac{k_z}{k}\sin\varphi \mathbf{y} - \frac{\varepsilon_t}{\varepsilon_z}\frac{k_\rho}{k}\mathbf{z}, \tag{62}$$

$$\hat{\varphi}(\mathbf{k}) = -\sin\varphi \mathbf{x} + \cos\varphi \mathbf{y}, \tag{63}$$

$$\hat{k}(\mathbf{k}) = \frac{k_\rho}{k}\cos\varphi \mathbf{x} + \frac{k_\rho}{k}\sin\varphi \mathbf{y} + \frac{k_z}{k}\mathbf{z}. \tag{64}$$

From Eq. (62) – Eq. (64), the essential integrals for Eq. (59) – Eq. (61) are,

$$\int_0^{2\pi} \cos\varphi \cdot e^{ik_\rho \rho \cos(\varphi-\alpha)} e^{im\varphi} d\varphi$$
$$= \pi i^{m+1} \cdot J_{m+1}(k_\rho \rho) e^{i(m+1)\alpha} + \pi i^{m-1} \cdot J_{m-1}(k_\rho \rho) e^{i(m-1)\alpha}, \tag{65}$$

$$\int_0^{2\pi} \sin\varphi \cdot e^{ik_\rho \rho \cos(\varphi-\alpha)} e^{im\varphi} d\varphi$$
$$= \pi i^m \cdot J_{m+1}(k_\rho \rho) e^{i(m+1)\alpha} + \pi i^{m-2} \cdot J_{m-1}(k_\rho \rho) e^{i(m-1)\alpha}, \tag{66}$$

$$\int_0^{2\pi} e^{ik_\rho \rho \cos(\varphi-\alpha)} e^{im\varphi} d\varphi = 2\pi i^m \cdot J_m(k_\rho \rho) e^{im\alpha}. \tag{67}$$

In Eq. (65) – Eq. (67), $J_m(z)$ is the $m^{th}$ order Bessel function of the first kind. By combining Eq. (59) – Eq. (67), we achieve the aimed reduction in the fold of integration.

APPENDIX II. DERIVATIONS FOR EQS. (53) - (54) AND EQS. (55) - (56) IN SECTION III.B

In this Appendix, within the SRM, we derive the wave matrix **Q** for a spherical and a planar interface. It is remembered from Eqs. (12) and (13) that for both cases the physical quantities to be matched at an interface are,

$$\mathbf{E}^{\|} + d_\perp \nabla_{\|} E^\perp, \tag{68}$$

$$\mathbf{H}^{\|} + i\omega d_{\|} \mathbf{n} \times \mathbf{D}^{\|}. \tag{69}$$

For the spherical case, we expand the fields in a region as,

$$\mathbf{E}(\mathbf{r}) = \sum_{nm} \left[ \mathbf{M}_{nm}(k,\mathbf{r}) \cdot a_{nm} + \mathbf{N}_{nm}(k,\mathbf{r}) \cdot b_{nm} \right], \tag{70}$$

$$\mathbf{H}(\mathbf{r}) = \sum_{nm} \frac{1}{iZ} \cdot \left[ \mathbf{N}_{nm}(k,\mathbf{r}) \cdot a_{nm} + \mathbf{M}_{nm}(k,\mathbf{r}) \cdot b_{nm} \right]. \tag{71}$$

In Eq. (70) and Eq. (71), $k$ and $Z$ are the wavenumber and the wave impedance of the region. The **M** and the **N** functions are defined as in [61],

$$\mathbf{M}_{nm}(k,\mathbf{r}) = z_n(kr) \cdot \mathbf{X}_{nm}(\theta,\varphi), \tag{72}$$

$$\mathbf{N}_{nm}(k,\mathbf{r}) = n(n+1) \frac{z_n(kr)}{kr} \cdot Y_{nm}(\theta,\varphi) \hat{\mathbf{r}} + \frac{1}{kr} \frac{\partial(rz_n(kr))}{\partial r} \cdot \mathbf{Z}_{nm}(\theta,\varphi). \tag{73}$$

In Eq. (72) and Eq. (73), $Y_{nm}$ and $\mathbf{X}_{nm}$, $\mathbf{Z}_{nm}$ are known as the scalar and vector spherical harmonics. The definition for the former can be found in (e.g., Appendix D of [61]) and will not be repeated here, while the ones for the latter are,

$$\mathbf{X}_{nm}(\theta,\varphi) = \frac{1}{\sin\theta} \frac{\partial Y_{nm}(\theta,\varphi)}{\partial \varphi} \hat{\boldsymbol{\theta}} - \frac{\partial Y_{nm}(\theta,\varphi)}{\partial \theta} \hat{\boldsymbol{\varphi}}, \tag{74}$$

$$\mathbf{Z}_{nm}(\theta,\varphi) = \frac{\partial Y_{nm}(\theta,\varphi)}{\partial \theta} \hat{\boldsymbol{\theta}} + \frac{1}{\sin\theta} \frac{\partial Y_{nm}(\theta,\varphi)}{\partial \varphi} \hat{\boldsymbol{\varphi}}. \tag{75}$$

$\mathbf{X}_{nm}$ and $\mathbf{Z}_{nm}$ hold the following orthogonality properties,

$$\int_0^{2\pi}\int_0^\pi \mathbf{X}^*_{n'm'}(\theta,\varphi) \cdot \mathbf{X}_{nm}(\theta,\varphi) \sin\theta d\theta d\varphi = n(n+1)\delta_{nn'}\delta_{mm'}, \tag{76}$$

$$\int_0^{2\pi}\int_0^\pi \mathbf{Z}^*_{n'm'}(\theta,\varphi) \cdot \mathbf{Z}_{nm}(\theta,\varphi) \sin\theta d\theta d\varphi = n(n+1)\delta_{nn'}\delta_{mm'}, \tag{77}$$

$$\int_0^{2\pi}\int_0^\pi \mathbf{X}^*_{n'm'}(\theta,\varphi) \cdot \mathbf{Z}_{nm}(\theta,\varphi) \sin\theta d\theta d\varphi = 0, \tag{78}$$

$$\int_0^{2\pi}\int_0^\pi \mathbf{Z}^*_{n'm'}(\theta,\varphi) \cdot \mathbf{X}_{nm}(\theta,\varphi) \sin\theta d\theta d\varphi = 0. \tag{79}$$

In Eq. (76) – Eq. (79), the integrations are done with respect to all elevation and azimuthal angles.

By using the expansions in Eq. (70) and Eq. (71) and the **M** and the **N** functions in Eq. (72) and Eq. (73), we express the physical quantities in Eq. (68) and Eq. (69) to be matched at the interface as,

$$\mathbf{E}^{\|} + d_\perp \nabla_{\|} E^\perp$$
$$= \sum_{nm} \left[ z_n(kr) \cdot \mathbf{X}_{nm}(\theta,\varphi) \cdot a_{nm} \right]$$
$$+ \sum_{nm} \left[ \frac{1}{kr} \frac{\partial(rz_n(kr))}{\partial r} \cdot \mathbf{Z}_{nm}(\theta,\varphi) \cdot b_{nm} \right] \tag{80}$$
$$+ d_\perp \cdot \sum_{nm} \left[ n(n+1) \frac{z_n(kr)}{kr^2} \cdot \mathbf{Z}_{nm}(\theta,\varphi) \cdot b_{nm} \right],$$

$$\mathbf{H}^{\|} + i\omega d_{\|} \mathbf{n} \times \mathbf{D}^{\|}$$
$$= \sum_{nm} \frac{1}{iZ} \cdot \left[ \frac{1}{kr} \frac{\partial(rz_n(kr))}{\partial r} \cdot \mathbf{Z}_{nm}(\theta,\varphi) \cdot a_{nm} \right]$$
$$+ \sum_{nm} \frac{1}{iZ} \cdot \left[ z_n(kr) \cdot \mathbf{X}_{nm}(\theta,\varphi) \cdot b_{nm} \right] \tag{81}$$
$$+ i\omega d_{\|} \cdot \varepsilon \cdot \sum_{nm} \left[ z_n(kr) \cdot \mathbf{Z}_{nm}(\theta,\varphi) \cdot a_{nm} \right]$$
$$- i\omega d_{\|} \cdot \varepsilon \cdot \sum_{nm} \left[ \frac{1}{kr} \frac{\partial(rz_n(kr))}{\partial r} \cdot \mathbf{X}_{nm}(\theta,\varphi) \cdot b_{nm} \right].$$

Next, we apply the orthogonality properties in Eqs. (76) –(79), so that Eqs. (80) and (81) split into two systems: (I) a TE system, which corresponds to the **M** function and $a_{nm}$ in Eqs. (70) and (71),

$$\mathbf{Q} = \begin{pmatrix} \int_\Omega \mathbf{X}^*_{nm}(\theta,\varphi) \cdot \left(\mathbf{E}^{\|} + d_\perp \nabla_{\|} E^\perp\right) d\Omega \\ \int_\Omega \mathbf{Z}^*_{nm}(\theta,\varphi) \cdot \left(\mathbf{H}^{\|} + i\omega d_{\|} \mathbf{n} \times \mathbf{D}^{\|}\right) d\Omega \end{pmatrix}$$
$$= n(n+1) \cdot \begin{pmatrix} z_n(kr) \\ \frac{1}{iZ} \cdot \frac{1}{kr} \frac{\partial(rz_n(kr))}{\partial r} + i\omega d_{\|} \cdot \varepsilon \cdot z_n(kr) \end{pmatrix} \cdot a_{nm}, \tag{82}$$

(II) a TM system, that corresponds to the **N** function and $b_{nm}$ in Eqs. (70) and (71),

$$\mathbf{Q} = \begin{pmatrix} \int_\Omega \mathbf{Z}^*_{nm}(\theta,\varphi) \cdot \left(\mathbf{E}^{\|} + d_\perp \nabla_{\|} E^\perp\right) d\Omega \\ \int_\Omega \mathbf{X}^*_{nm}(\theta,\varphi) \cdot \left(\mathbf{H}^{\|} + i\omega d_{\|} \mathbf{n} \times \mathbf{D}^{\|}\right) d\Omega \end{pmatrix}$$
$$= n(n+1) \cdot \begin{pmatrix} \frac{1}{kr}\left[\frac{\partial(rz_n(kr))}{\partial r} + d_\perp \cdot n(n+1) \cdot \frac{z_n(kr)}{r}\right] \\ \frac{1}{iZ} \cdot z_n(kr) + i\omega d_{\|} \cdot \varepsilon \cdot -\frac{1}{kr} \frac{\partial(rz_n(kr))}{\partial r} \end{pmatrix} \cdot b_{nm}. \tag{83}$$

Eq. (82) and Eq. (83) give the **Q** matrix in Eqs. (53) and (54) in the main text.

For the planar case, we focus on the TE wave in the region,

$$\mathbf{E}(\mathbf{r}) = \hat{\varphi}(\mathbf{k}) e^{i\mathbf{k}_{\|} \cdot \mathbf{r}_{\|} + ik_z z} \cdot a(k_z), \tag{84}$$

$$\mathbf{H}(\mathbf{r}) = -\frac{k}{\omega\mu_0} \hat{\theta}(\mathbf{k}) e^{i\mathbf{k}_{\|} \cdot \mathbf{r}_{\|} + ik_z z} \cdot a(k_z). \tag{85}$$

In Eqs. (84) and (85), we assume the wave propagates along a wave vector **k**. $\hat{\varphi}$, $\hat{\theta}$ and $\hat{k}$ (which is the unit vector along the wave vector **k**) form a right-handed system,

$$\hat{\varphi}(\mathbf{k}) = \mathbf{X}(\mathbf{k}_{\|}) = -\frac{k_y}{k_\rho}\mathbf{x} + \frac{k_x}{k_\rho}\mathbf{y}, \tag{86}$$

$$\hat{\theta}(\mathbf{k}) = \frac{k_z}{k} \mathbf{Z}(\mathbf{k}_\parallel) - \frac{k_\rho}{k} \mathbf{z} = \frac{k_z}{k}\left(\frac{k_x}{k_\rho}\mathbf{x} + \frac{k_y}{k_\rho}\mathbf{y}\right) - \frac{k_\rho}{k}\mathbf{z}. \quad (87)$$

In the above, $\mathbf{k}$, $\mathbf{k}_\parallel$, $k_x$, $k_y$, $k_z$, $k_\rho$ and $k$ are introduced in the same way as in Eqs. (24) and (25) in the main text, and two functions $\mathbf{X}$ and $\mathbf{Z}$ are defined for later use. We substitute Eqs. (84) and (85) in Eqs. (68) and (69) for the physical quantities to be matched at the interface,

$$\mathbf{E}^\parallel + d_\perp \nabla_\parallel E^\perp = \mathbf{X}(\mathbf{k}_\parallel) e^{i\mathbf{k}_\parallel \cdot \mathbf{r}_\parallel + ik_z z}, \quad (88)$$

$$\mathbf{H}^\parallel + i\omega d_\parallel \mathbf{z} \times \mathbf{D}^\parallel = \left(-\frac{k_z}{\omega\mu_0} - i\omega d_\parallel \varepsilon_0 \varepsilon_t\right)\mathbf{Z}(\mathbf{k}_\parallel) e^{i\mathbf{k}_\parallel \cdot \mathbf{r}_\parallel + ik_z z}. \quad (89)$$

We project them onto the $\mathbf{X}$ and $\mathbf{Z}$ functions and get the wave matrix,

$$\mathbf{Q} = \begin{pmatrix} \mathbf{X}(\mathbf{k}_\parallel) \cdot (\mathbf{E}^\parallel + d_\perp \nabla_\parallel E^\perp) \\ \mathbf{Z}(\mathbf{k}_\parallel) \cdot (\mathbf{H}^\parallel + i\omega d_\parallel \mathbf{z} \times \mathbf{D}^\parallel) \end{pmatrix}$$
$$= \begin{pmatrix} 1 \\ -\dfrac{k_z}{\omega\mu_0} - i\omega d_\parallel \cdot \varepsilon_0 \varepsilon_t \end{pmatrix} e^{i\mathbf{k}_\parallel \cdot \mathbf{r}_\parallel + ik_z z}. \quad (90)$$

Then, we shift to the TM wave in the region,

$$\mathbf{E}(\mathbf{r}) = \left[\frac{q_z}{q}\mathbf{Z}(\mathbf{k}_\parallel) - \frac{\varepsilon_t}{\varepsilon_z}\frac{k_\rho}{q}\mathbf{z}\right]e^{i\mathbf{k}_\parallel \cdot \mathbf{r}_\parallel + iq_z z} \cdot b(q_z), \quad (91)$$

$$\mathbf{H}(\mathbf{r}) = \frac{\omega\varepsilon_0\varepsilon_t}{q}\mathbf{X}(\mathbf{k}_\parallel) e^{i\mathbf{k}_\parallel \cdot \mathbf{r}_\parallel + iq_z z} \cdot b(q_z). \quad (92)$$

In Eqs. (91) and (92), $q_z$ is the $z$ component of the wave vector of the TM wave. By following the same procedures as in Eq. (88) – Eq. (90), we obtain the wave matrix,

$$\mathbf{Q} = \begin{pmatrix} \mathbf{Z}(k_x, k_y) \cdot (\mathbf{E}^\parallel + d_\perp \nabla_\parallel E^\perp) \\ \mathbf{X}(k_x, k_y) \cdot (\mathbf{H}^\parallel + i\omega d_\parallel \hat{\mathbf{n}} \times \mathbf{D}^\parallel) \end{pmatrix}$$
$$= \begin{pmatrix} \dfrac{q_z}{q} - \dfrac{\varepsilon_t}{\varepsilon_z}\dfrac{ik_\rho^2 d_\perp}{q} \\ \dfrac{\omega\varepsilon_0\varepsilon_t}{q}(1 + id_\parallel q_z) \end{pmatrix} \cdot e^{i\mathbf{k}_\parallel \cdot \mathbf{r}_\parallel + iq_z z}. \quad (93)$$

By replacing $z$ in Eqs. (90) and (93) with a position relative to the interface (which is located at $z_0$), i.e., $z - z_0$, we obtain Eqs. (55) and (56) in the main text.